\documentclass
[twocolumn,preprintnumbers,amsmath,amssymb,showpacs,prb,aps]{revtex4}
\usepackage{txfonts}
\usepackage{pifont}
\usepackage{amssymb}
\usepackage{dcolumn}
\usepackage{amsmath}
\usepackage[dvips]{epsfig}

\makeatletter
\def\btt#1{\texttt{\@backslashchar#1}}%
\DeclareRobustCommand\bblash{\btt{\@backslashchar}}%
\makeatother

\begin{document}
\title{Influence of electronic correlations on orbital polarizations in the
       parent and doped iron pnictides}
\author{Ya-Min Quan$^{1}$, Liang-Jian Zou$^{1,2 \footnote{Correspondence author,
        Electronic mail: zou@theory.issp.ac.cn}}$, Da-Yong Liu$^{1}$,
        and Hai-Qing Lin$^{2}$}
\affiliation{ \it $^1$ Key Laboratory of Materials Physics,
              Institute of Solid State Physics, Chinese Academy of Sciences,
               P. O. Box 1129, Hefei 230031, China\\
         \it $^2$ Department of Physics, Chinese University of Hong Kong,
                  Shatin, New Territory, Hong Kong, China\\}
\date{Nov 6, 2010}

\begin{abstract}
Orbital polarization and electronic correlation are two essential
aspects in understanding the normal state and superconducting
properties of multiorbital FeAs-based superconductors. In this Letter, we
present a systematical study on the orbital polarization of iron
pnictides from weak to strong Coulomb correlations within the KRSB
approach. The magnetic phase diagram of the two-orbital model for
LaFeAsO clearly shows that the striped antiferromagnetic metallic
phase with considerable orbital polarization exists over a wide
doping range. The reversal of the orbital polarization occurs in the
intermediate correlation regime.A small crystal field splitting
enhances the orbital density wave order.
\end{abstract}

\maketitle

Since the discovery of high-$T_{c}$ superconductivity (SC) in doped
iron pnictides \cite{Hosono,XHChen,WangNL}, the iron-based
high-$T_c$ SC material has been the subject of intensive researches
in the past two years. Among various normal-state and SC properties
of iron pnictides, the roles of the electronic correlation and the
orbital polarization have been the central topics. The parent phases
of cuprates are Heisenberg antiferromagnets (AFM) and insulators,
and the important role of the electronic correlation and dominant
single-orbital hole with the $x^{2}-y^{2}$ symmetry were well
established and verified. Unlike the high-$T_{c}$ cuprates,
the parent phase of iron
pnictide SC are a bad metal\cite{XHChen}, and usually exhibits
spin-density-wave (SDW)-type or striped AFM ordering
\cite{WangNL,DaiPC}. What is the role of the electronic correlation
and how many Fe 3\emph{d} orbits are involved in the iron pnictides
are two essential and key aspects to understand various properties
in the SDW metallic and SC phase.

Up to date, experimental evidences on the role of electronic
correlation in parent phase and SC phases are controversial. Indeed,
the metallic conductivity \cite{Hosono,XHChen}, small magnetic
moment of Fe spins \cite{DaiPC}, X-ray absorption spectroscopy (XAS)
and resonant inelastic X-ray scattering (RIXS) of Fe \cite{WLYang}
in undoped iron pnictides seem to suggest a weak correlation between
Fe 3\emph{d} electrons. From the resonant X-ray emission spectroscopy
(RXES) of Fe 3\emph{d} $L_{{2,3}}$ edges, Kurmaev et al.
\cite{EZKurmaev} suggested that the $111$ phase is a weakly or at
most moderately correlated system. However, many more experimental
clues demonstrated the band narrowing effect, suggesting the
important roles of the 3\emph{d} electronic correlation
\cite{WLYang,EZKurmaev,HDing}. In fact, in addition to the band
narrowing effect, the existence of intermediate magnetic moment in
the $1111$ , $122$ and $111$ phases also shows the importance of the
electronic correlation. Therefore one should employ an approach
applicable for both weak and strong Coulomb interactions so as to
uncover the roles of electronic correlations on the electronic states
and magnetic configuration in iron pnictides.

On the other hand, it is still under hot debate in literature on how
many orbits are involved and what the orbital symmetry of the bands
near $E_{F}$ is in the parent and doped iron pnictides.
Theoretically, two-orbital \cite{SCZhang}, three-orbital
\cite{PALee}, four-orbital \cite{Korshunov} and even five-orbital
\cite{Haule,Kuroki} tight-binding model were proposed. Most of the
multi-orbital tight-binding models captured the major Fermi surface
and dominant band structures  characters near $E_{{F}}$ of the
$1111$ and $122$ phases. To distinguish the validity of these models
is essential to reveal the symmetry character and the number of the
orbits involved in various iron pnictides.

The role of the electronic correlation in the groundstate properties
of parent iron pnictides was discussed in very recent papers
\cite{zhou,ko}. To uncover the roles of the electronic correlation
and orbital polarization in high-Tc iron pnictide SC, we apply the
Kotliar-Ruckenstein's slave boson (KRSB) method \cite{Kotliar} on
the two-orbital Hubbard models. The KRSB approach and its extension
may provide a useful tool to treat the multi-orbital Hubbard model
over a wide correlation range. It also has a few of advantages in
dealing with multiple orbitals, nontrivial magnetic configurations
and spatial correlations. Our recent numerical ansatz developed for
the multi-orbital KRSB solution overcomes the convergency of many
parameters in minimizing the groundstate energy, suitable for
treating arbitrary coulomb interaction in the presence of
interorbital hoppings or hybridizations and various magnetic
configurations.

In this paper, we first present the magnetic phase diagram of
two-orbital model and demonstrate that in the intermediate
electronic correlation, the striped AFM (SAFM) phase with small
magnetic moment is stable against the paramagnetic (PM),
ferromagnetic (FM) and {\it N\'{e}el} AFM phases, accompanied with a
small orbital polarization, or a ferro-orbital density wave (F-ODW)
order. A small crystal field splitting arising from the lattice
distortion further stabilizes the SAFM and F-ODW ground state. The
evolution of the orbital occupation and Fermi surface with the
electron correlation shows that the contribution of the $d_{xz}$
orbital symmetry is larger than that of the $d_{yz}$ orbital. These
results clearly show the importance of the intermediate electronic
correlation in the parent and doped phases of iron pnictides.


We start with two-orbital model
Hamiltonian $H=H_{0}+H_{I}$ applied for iron pnictides in real space
with
\begin{eqnarray}
  H_{0}&=&-\sum_{<ij>\alpha\beta\sigma}\left(t_{\alpha\beta}c^{\dagger}_{i\alpha
  \sigma}c_{j\beta\sigma}+h.c.\right)+ \sum_{i\alpha\sigma}\left(\varepsilon_{
  \alpha}-\mu\right)n_{i\alpha\sigma}  \\
  H_{I}&=&U\sum_{i\alpha}n_{i\alpha\uparrow}n_{i\alpha\downarrow}
  +\sum_{i\sigma\sigma^{\prime}\left(
  \alpha>\beta\right)}
   \left(U^{\prime}-J_{H}\delta_{\sigma\sigma^{\prime}}\right)n_{i\alpha\sigma}
   n_{i\beta\sigma^{\prime}}               \nonumber\\
   &-&J_{H}\sum_{i\alpha\neq\beta}\left(c^{\dagger}_{i\alpha\uparrow}c_{i\alpha
   \downarrow}c^{\dagger}_{i\beta\downarrow}c_{i\beta\uparrow}
   -c^{\dagger}_{i\alpha\uparrow}c^{\dagger}_{i\alpha\downarrow}c_{i\beta
   \downarrow}c_{i\beta\uparrow}\right),
\end{eqnarray}
where $C_{i\alpha\sigma}^{\dagger}$ creates an electron with orbital
index $\alpha$ and spin $\sigma$ at lattice site $i$,
$n_{i\alpha\sigma}$ is the corresponding occupation number operator.
The hopping integral for the orbital $\alpha$ and $\beta$ is denoted
by $t_{\alpha\beta}$. The intraband (inter-band) Coulomb repulsion
and Hund's rule coupling are denoted by $U$ ($U^{\prime}$) and
$J_{H}$ , respectively. Here we set $U^{\prime}=U-2J_{H}$.

To reflect the multi-orbital character of iron pnictides, we extend
the single-orbital KRSB approach \cite{Kotliar} to the two-orbital
Hubbard models for various magnetic configurations. In the
multiorbital Hubbard model, a few of auxiliary boson field operators
representing the possibilities of various electron occupations are
introduced, such as $e, p, d, b, t, q,$ which denote the possibilities
of none, single, double, triplicate, quaternity occupations. With
these auxiliary boson fields, an original fermion operator can be
expressed as:
\begin{eqnarray}
  c^{\dag}_{i\alpha\sigma}&=&Q^{-\frac{1}{2}}_{i\alpha\sigma}
  \left(p^{\dag}_{i\alpha\sigma}e_{i}+
  b^{\dag}_{i\alpha}p_{i\alpha\sigma}
  +\sum_{\sigma'}d^{\dag}_{i\sigma_{\alpha}\sigma'_{\beta}}p_{i\beta\sigma'}
  +t^{\dag}_{i\alpha\sigma}b_{i\beta}\right. \nonumber\\
  &+& \left.\sum_{\sigma'}
  t^{\dag}_{i\beta\sigma'}d_{i\bar{\sigma}_{\alpha}\sigma'_{\beta}}
  +q^{\dag}_{i}t_{i\alpha\bar{\sigma}}\right)
  (1-Q_{i\alpha\sigma})^{-\frac{1}{2}}f^{\dag}_{i\alpha\sigma}
\end{eqnarray}
Here $f^{\dag}_{i\alpha\sigma}$ is the new slaved fermion operator
and $Q_{i\alpha\sigma}$ is an auxiliary particle number operator
\cite{Kotliar}. Projecting the original fermion operators into these
boson field and fermion field operators, one could not only obtain
an effective Hamiltonian, but also get the groundstate energy in the
saddle point approximation with the normalization condition and the
fermion number constraints \cite{Kotliar}. Here we employ a
generalized Lagrange multiplier method to enforce these constraint
conditions, thus the interorbital hoppings and crystal field
splitting can be treated on the same foot. The fermion occupation
number is constrained with the penalty function method.
To enforce the normalization condition, we have a boundary
constrained condition:
\begin{eqnarray}\label{eq:boundary constraint}
 1 &\geq &\sum_{\alpha\sigma}p_{\alpha\sigma}^{2}+\sum_{\alpha}b_{\alpha}^{2}
 +\sum_{\alpha\sigma\sigma^{\prime}}
 d_{\alpha\sigma\sigma^{\prime}}^{2}+
 \sum_{\alpha\sigma}t_{\alpha\sigma}^{2}+q^{2}
\end{eqnarray}
We use optimizing method to get the minimized groundstate energy. To
search for possible ground state, we study four different magnetic
configurations, including the PM or nonmagnetic, FM, {\it N\'{e}el}
AFM and SAFM  phases so as to find the stable ground state.

 \begin{figure}[htbp]
 \centering
\includegraphics[angle=0, width=0.9 \columnwidth]{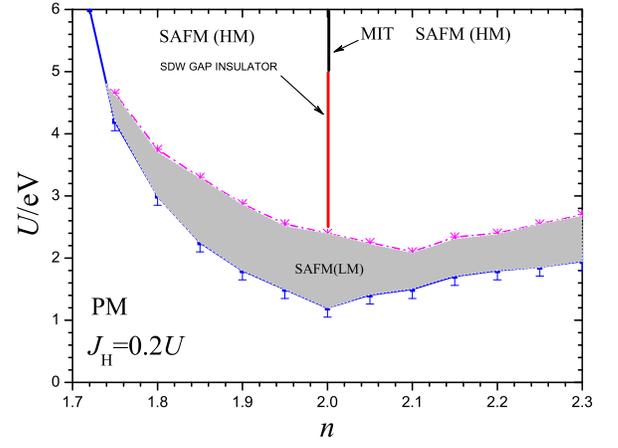}
\caption{Magnetic phase diagram as a function of particle density
$n$ and correlated $U$ for $J_{H}=0.2U$. PM, SAFM(LM) and SAFM(HM) stand for
paramagnetic metallic, striped antiferromagnetic metallic phases
with magnetic moment $m<1 \mu_{B}$ and $m>1
\mu_{B}$, respectively.}
\label{fig:Diagram}
\end{figure}

As the penalty factor is sufficiently large, the particle number
constraint is satisfied. Nevertheless, the problem of boundary
constrained condition is still difficult. In numerically searching
the global minima of the groundstate energy, we employ the pattern
search method, together with the gradient method and the Rosenbrock
method. If the optimizing point is on the boundary, we move one step
inward the high-dimensional ellipsoid and the equipotential plane.
Since the first axis of the new local orthogonal coordinate system
of Rosenbrock method directs to the negative gradient direction, the
corresponding algorithm is simple.
To investigate the groundstate electronic and magnetic properties of
iron pnictides, we adopt the hopping parameters of Zhang {\it et
al}. \cite{SCZhang} for the two-orbital situation in the present
slave boson scheme. In the two-orbital situation, the orbits $1$ and $2$
refer to the $d_xz$ and $d_yz$ components, respectively.


We first obtain the magnetic-phase diagram of the two-orbital system
through comparing the energies of the PM (or nonmagnetic), FM, {\it
N\'{e}el} AFM and SAFM configurations. In the present symmetric
orbital situation, the penalty factors adopted for the two
degenerate orbits are identical. This will save a great deal of
computation time in minimizing the total energy.

\begin{figure}[htbp]
\centering
\includegraphics[angle=0, width=0.9 \columnwidth]{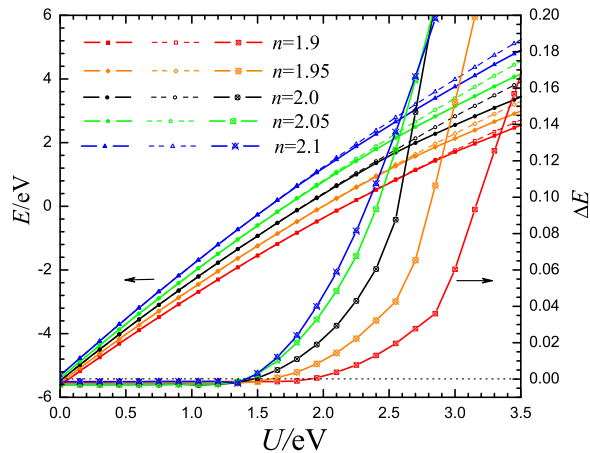}
\caption{Electronic correlation dependence of total energies of PM
phase (dashed line) and SAFM phase (solid line) and that of the energy
difference ($E_{PM}$-$E_{SAFM}$) between these two magnetic
configurations at $n=1.9, 1.95, 2, 2.05$ and $2.10$, respectively.}
\label{fig:Energy}
\end{figure}

Fig.\ref{fig:Diagram} displays the zero-temperature phase diagram of
$U$ vs $n$ for Zhang {\it et al.}'s\cite{SCZhang} hopping
parameters. We consider the full Hund's coupling $J_{H}$, which
equals to $0.2U$. One notices that only the PM metallic phase is
stable when the on-site Coulomb correlation $U<U_{c1}\approx 1.2$
eV. When $U>U_{c1}$, the SAFM phases are stable over wide doping and
interaction ranges. This SAFM region can be divided into high
magnetic moment one (SAFM(HM)) with $m>1 \mu_{B}$ and low magnetic
moment one (SAFM(LM)) with $m<1 \mu_{B}$. The latter lying in
$U_{c1}<U<U_{c2}$ is the most interesting and marked with a shadow
since the magnetic moments of the parent and doped phases of the most of iron
pnictides, including the 1111, 122 and 111 phases, fall into this
region. In the present phase diagram, it is also shown that the
SAFM(LM) regime is limited to a very narrow range.

Especial interest is found around the particle filling number of
$n=2$. One notices that the phase diagram is not symmetrical about
the half filling ($n=2$) due to the presence of the next
nearest-neighbor hopping. When $U>U_{c1}$, the system enters the
spin gapped insulating phase. In this insulating region the gap
between the spin-down and spin-up subbands, the sublattice magnetic
moment, the spin exchange splitting and the quasiparticle bandwidth
increase with the increase of $U$.
When the electronic correlation becomes so strong that $U>5$ eV, the
system enters a full gapped region, very similar to the conventional
Mott insulating phase. In the insulating region, with the further
increase of the Coulomb interaction $U$, the spin exchange splitting
approaches a constant and sublattice magnetic moment is saturated.
In comparison with the conventional PM Mott phase, the quasiparticle
bandwidth in this SAFM Mott phase gradually decrease with the
increase of $U$ again.
As shown in the following, the orbital polarization is observed in
a wide doping and an interaction regime, while no orbital
polarization is seen at $n=2$ when the correlation is strong.

\begin{figure}[thbp]
\centering
\includegraphics[angle=0, width=0.9 \columnwidth]{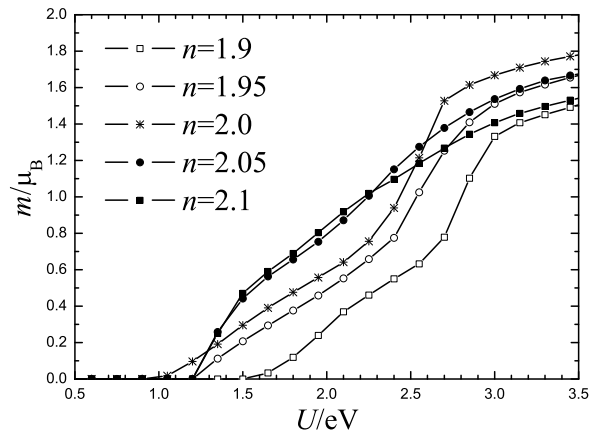}
\caption{Sublattice magnetic moment in the SAFM phase as a function
of the Coulomb repulsion strength $U$. Theoretical parameters are
the same to Fig.1.} \label{fig:Magnetism}
\end{figure}

For more clarity, we plot the correlation dependence of the
groundstate energies in the PM structure and the SAFM one in
Fig.\ref{fig:Energy}. Also an energy difference, $E_{{\emph
PM}}-E_{\emph{SAFM}}$, between these two phases is displayed. One
finds that when the electronic correlation is weak, or
$U<U_{\emph{c1}}$, the energy difference is negative and very small.
So the PM phase is stable against the SAFM one. When the correlation
reaches to the intermediate strength that $U>U_{\emph{c1}}$, the
energy difference becomes positive, the SAFM phase becomes more
stable than the PM phase. Furthermore, one could estimate the
magnitude of the magnetic couplings between iron spins. The magnetic
coupling strength crucially depends on the doping concentration and
electron correlations, as seen in Fig.\ref{fig:Energy}. For a
typical parent compound LaOFeAs, the on-site Coulomb correlation
between Fe $3d$ electrons is estimated to $2\sim3$ eV
\cite{Anisimov}. This gives rise to the magnetic coupling strength
about 20 meV at $n=2$, in the same magnitude order of the spin
coupling between Fe ions in LaOFeAs deduced from the neutron
scattering experiment \cite{DaiPC}.

Fig.\ref{fig:Magnetism} shows the evolution of sublattice magnetic
moments with the increase of the electronic correlation in the SAFM
phase. It shows that the system remains PM without any magnetic
ordering unless $U$ becomes larger than $U_{\emph{c1}}$, which
increases with hole doping. In the realistic parameter range, {\it
i.e.}, intermediate correlation regime, the magnetization decreases
with the increase of hole doping, in agreement with the experiment
observation.
We notice that the minimal value of $U_{\emph{c1}}$ is about $1.2$
eV at $n=2$, which is considerably larger than the mean-field
approximation results for the two-orbital model
\cite{Rong,Daghofer}. It is also found that the sublattice magnetic
moment increases continuously until $U$ reaches $U_{\emph{c2}}=2.4$
eV, where it changes discontinuously. This behavior is also found by
Yu {\it et al}. in the mean-field solution in the four-orbital model
\cite{Rong}. Further, the upper critical value for the SAFM regime
with $m<1 \mu_{B}$ is $2.4$ eV for $n=2.0$, significantly larger
than that obtained by the mean-field approximation
\cite{Rong,Daghofer}, suggesting that the quantum fluctuations are
well considered within the present KRSB approach. At $n=2.0$, the
sublattice magnetic moment is larger than $1$ $\mu_{B}$ when
$U>U_{c2}=2.4$ eV. The SAFM system enters the high magnetic moment
region, this is also shown in Fig.\ref{fig:Diagram}. By monitoring
the density of states (DOS) near the Fermi surface, we find that due
to strong correlation, the system becomes insulating at $n=2.0$.
%
%
We notice such a fact that when the systems cross over from the SAFM
insulating phase to the full-gap Mott one, the sublattice magnetic
moments of the systems with $n=1.9$ and $2.1$, or $n=1.95$ and $2.05$,
approach the same values, as seen in Fig.3, showing that the
electron-hole symmetry restores in the strong correlation regime.

\begin{figure}[htbp]
\centering
\includegraphics[angle=0, width=0.9 \columnwidth]{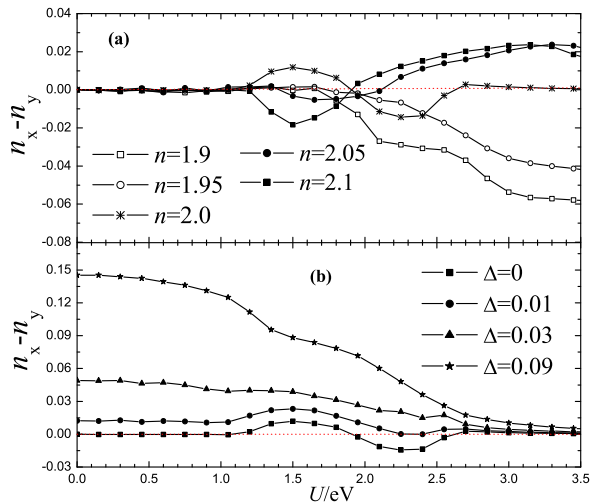}
\caption{Magnetic and crystal field dependence of orbital
polarizations or itinerant orbital density wave ordering in
two-orbital models. Theoretical parameters are the same to Fig.1.}
\label{fig:orbital order}
\end{figure}
\begin{figure}[thbp]
\centering
\includegraphics[angle=0, width=0.9 \columnwidth]{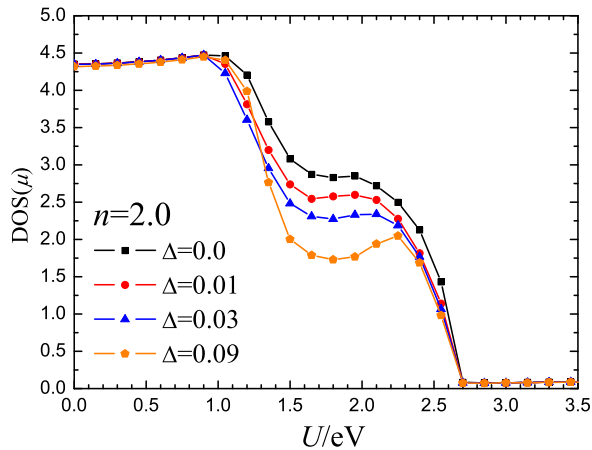}
\caption{The DOS at the Fermi surface as a function of $U$ at
different doping concentrations.}
 \label{fig:DOS2}
\end{figure}

In the present two-orbital model, we can address the orbital
polarization or orbital ordering problem in the parent and doped phases of
LaOFeAs. We find that in the weakly correlated PM phase , due to the
equivalence of the present two orbitals, the symmetry of two
orbitals is not broken, no orbital polarization or obrital ordering
is observed. As soon as the system enters the intermediate
correlated SAFM phase, the orbital polarization is observed, as seen
in Fig.4. Fig.4(a) shows that in the SAFM phase, the orbital
polarization considerably depends on the Coulomb correlation and
particle filling. When $n<2$, the orbital polarization is negative,
or the $d_{\emph{yz}}$ component is enhanced with the increase of
$U$. At $n=2$, the orbital polarization only occurs in the
intermediate correlation regime, and the polarization is reversed at
$U_{\circ}=1.9$ eV; when $U_{\emph{c1}}<U<U_{0}$ the
polarization is dominated by the $d_{xz}$ component, while it is
dominated by the $d_{\emph{yz}}$ component when
$U_{\circ}<U<U_{\emph{c2}}$. When $U>U_{\emph{c2}}$, due to strong
correlation, the two electrons occupy two orbitals with the same
possibility, giving rise to a vanishing polarization. This behavior
was also found in the conventional mean-field
approximation\cite{DXYAO}. When $n>2$, we notice that the
polarization also reverses the sign at about $U_{0}$. Though
such an orbital polarization of ordering is weak, it is greatly
enhanced in the presence of a small crystal field splitting, as
shown in Fig.4(b). From Fig.4(b) one finds that the orbital
polarization do not reverse over a wide correlation regime in the
presence of crystal field splitting, while a kink at $U=0$ still
exists for various splitting.

These results clearly demonstrate that the orbital polarization or
ferro-orbital ordering is accompanied with the onset of the SAFM
ordering in the intermediate correlated iron pnictides. Since in the
intermediate correlation regime, the system is metallic, a question
naturally arises that what the essential of the F-ODW ordering is in
metallic iron pnictides. Obviously, such an orbital polarization is
essentially itinerant, as the itinerant FM in transition metals.
Hence the orbital ordering in iron pnictides is characteristic of
density wave. As we pointed out in an early paper \cite{LuFeng},
such an itinerant ferro-orbital ordering can occur as soon as a
Stoner-like condition is satisfied \cite{LuFeng}. Besides,in
Fig.\ref{fig:DOS2},we also display the correlation dependence of the
quasiparticle DOS near $E_{\emph{F}}$ in the present two-orbital
systems. It is found that the DOS increases slowly in the weak
correlated PM region. With the increase of $U$,the DOS first steeply
decreases and then slightly increases in the metallic SAFM ordered
phase with small magnetic moment, demonstrating a pseudogap
behavior. The pseudogap in the intermediate correlation regime turns
to a full gap in the spin-density-wave gap insulating phase when
$U>2.7$ eV. From the analysis on the spin-dependent partial DOS, we
find this full gap is triggered by spin polarization. When $U > 4.1$
eV, the systen enters a conventional Mott insulator phase.


In summary, we have shown that in the parent and doped phases of iron
pnictides, the intermediate electronic correlation favors a striped
antiferromagnetic configuration accompanied with a small structural
distortion and weak ferro-orbital ordering. The orbital polarization
is the characteristics of a ferro-orbital density wave.
\\


This work was supported by the NSFC of China No.10874186,
11074257 and 11047154. Numerical calculations were
performed at the Center for Computational Science of CASHIPS.


\end{document}